# Neural fields for adaptive white matter microstructural modeling

Adam M. Saunders*[a], Gaurav Rudravaram[a], Elyssa M. McMaster[a], Michael E. Kim[b], Trent Schwartz[a], Yimeng Dou[a], Lianrui Zuo[a], Adam W. Anderson[c-e], Bennett A. Landman[a-e]
[a]Department of Electrical and Computer Engineering, Vanderbilt University, Nashville, TN, USA; [b]Department of Computer Science, Vanderbilt University, Nashville, TN, USA; [c]Department of Biomedical Engineering, Vanderbilt University, Nashville, TN, USA; [d]Department of Radiology and Radiological Sciences, Vanderbilt University Medical Center, Nashville, TN, USA; [e]Vanderbilt University Institute of Imaging Science, Vanderbilt University Medical Center, Nashville, TN, USA

## ABSTRACT

Diffusion magnetic resonance imaging (MRI) is sensitive to the micron-level diffusion motion of water molecules, allowing *in vivo* characterization of white matter architecture in the brain. Microstructural models like diffusion tensors or the Standard Model are typically fit independently on each voxel, limiting their ability to incorporate local spatial context and adapt to regional tissue complexity. Here, we introduce FiberField, an approach using neural fields that selects from multiple candidate models with different numbers of fibers to estimate fiber configurations and properties. This method incorporates spatial context through coordinate representations and model-weighted priors inspired by Bayesian model averaging to adapt to local microstructure. To compare FiberField with ground truth fiber properties, we validate on simulated diffusion MRI before applying to in vivo data. In simulated diffusion MRI, FiberField accurately estimates fiber directions and diffusivities using a multi-stick model (angular error: 2.2 ± 2.8°, diffusivity root mean square error (RMSE): $4.4 \times 10^{-4} \pm 5.2 \times 10^{-4}$ mm$^2$/s, fiber sensitivity: 0.87 ± 0.21). FiberField resolves fibers at 25% smaller angular separations than fod2fixel, and the diffusivity RMSE is 52% lower within fiber crossings than the Standard Model. On acquired diffusion MRI, FiberField's estimated fiber directions agree with those from MRtrix3's fod2fixel while also mapping microstructural properties such as diffusivity (angular difference: 10.5 ± 7.1°). Using neural fields allows for adaptive white matter modeling for selecting from a set of alternative models while maintaining accurate fiber property estimation.

## 1. INTRODUCTION

Changes in cerebral white matter fibers provide useful biomarkers to track aging, neurodegeneration, and disease status[1]. Diffusion magnetic resonance imaging (MRI) is sensitive to the random thermal motion of water molecules[2]. The diffusion of water molecules is highly constrained by the underlying fiber architecture, causing the measured diffusion signal to reflect tissue microstructure. Consequently, diffusion MRI provides a non-invasive means of probing white matter in the brain.

Many approaches for modeling white matter using diffusion MRI are inverse problems. One example is tractography, which is the process of reconstructing white matter fibers from diffusion MRI acquisition[3]. Oftentimes, tractography is a two-stage process that involves first estimating a voxel-wise microstructural model of fiber geometry (such as a fiber orientation distribution (FOD) calculated from constrained spherical deconvolution or CSD[4]), and then generating streamlines through the estimated principal directions of microstructural models[2]. However, this two-stage process is fundamentally interdependent; fiber properties cannot be estimated without knowing their direction, and accurate estimation of fiber directions requires knowledge of microstructural fiber properties.

This interdependence arises in both directions. First, fiber orientations impact microstructural modeling. Crossing white matter fibers, which occur in 60-90% of voxels[5], complicate microstructural modeling because multiple fiber populations contribute to the measured diffusion signal. Modeling each voxel using a single tensor can only account for a single fiber orientation, meaning microstructural metrics such as fractional anisotropy are artificially lower in crossings and are not specific[6], although multi-tensor models help mitigate these limitations[7,8]. While higher-order models such as FODs from CSD[4,9] or fiber orientation estimated using continuous axially symmetric tensors (FORECAST)[10] can represent multiple fibers, there are limits in angular resolution for distinguishing fibers depending on the maximum spherical harmonic order used to represent the FOD[4]. Even if the model can resolve multiple orientations, voxel-wise models typically assume that fiber properties do not change throughout a tissue type as in multi-shell multi-tissue CSD[9] or throughout a voxel as in the Standard Model of diffusion MRI and other similar multi-compartment models[11] (Fig. 1).

*adam.m.saunders@vanderbilt.edu

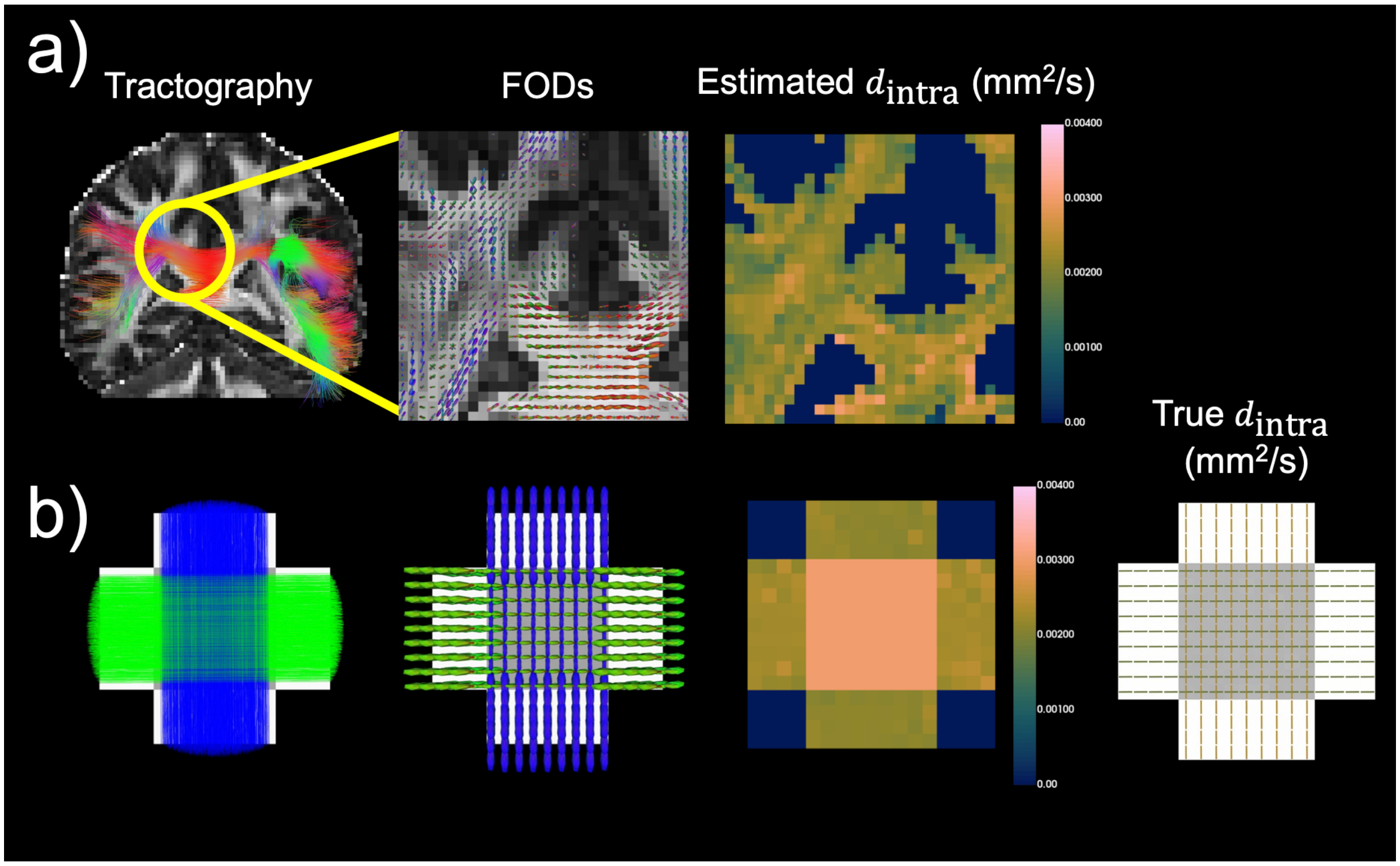

Figure 1. Traditional microstructural models for white matter in the brain rely on voxel-level mapping of diffusion properties, such as intra-axonal diffusivity from the Standard Model (a). However, fiber properties may differ in a voxel with fiber crossings, and Standard Model fitting can overestimate the intra-axonal diffusivity in crossings on simulated diffusion MRI (b).

On the other hand, microstructural modeling also impacts fiber estimation. Using diffusion tensors as the microstructural model for tractography can lead to inaccurate bundle estimation[12]. Even CSD fails to resolve crossings below about 40° angular separation[13]. There have been efforts to segment lobes of the resulting FOD into individual fibers[14,15] or unmix the effects of the underlying fibers[16]. The response function chosen for CSD can also impact tractography, leading to differences in stopping criteria for streamlines and undesired variability in the resulting streamlines[17].

There have been several other methods that attempt to model fibers and microstructural properties jointly. Convex optimization modeling for microstructural informed tractography (COMMIT) assumes fiber directions and numbers from candidate tractograms and fits microstructural properties based on these directions[3]. Forward modeling for complex microstructure (FORCE) simulates a dictionary of potential microstructural parameters and fiber geometries, and uses dictionary matching to find similar signals[18]. While these approaches couple orientation and microstructural modeling, they rely on predefined tractograms or simulated discrete dictionaries over a vast space of parameters, and they do not exploit local spatial context.

We introduce FiberField, drawing inspiration from deep learning frameworks that use implicit neural fields for fitting microstructural models such as CSD[19] and the Standard Model[20]. Implicit neural fields represent diffusion signals across space rather than fitting each voxel independently. By leveraging a coordinate representation, FiberField incorporates spatial context beyond independent voxelwise fitting while modeling fiber-specific properties. Additionally, FiberField adaptively selects from a set of candidate models based on the number of fibers in a voxel (Fig. 2).

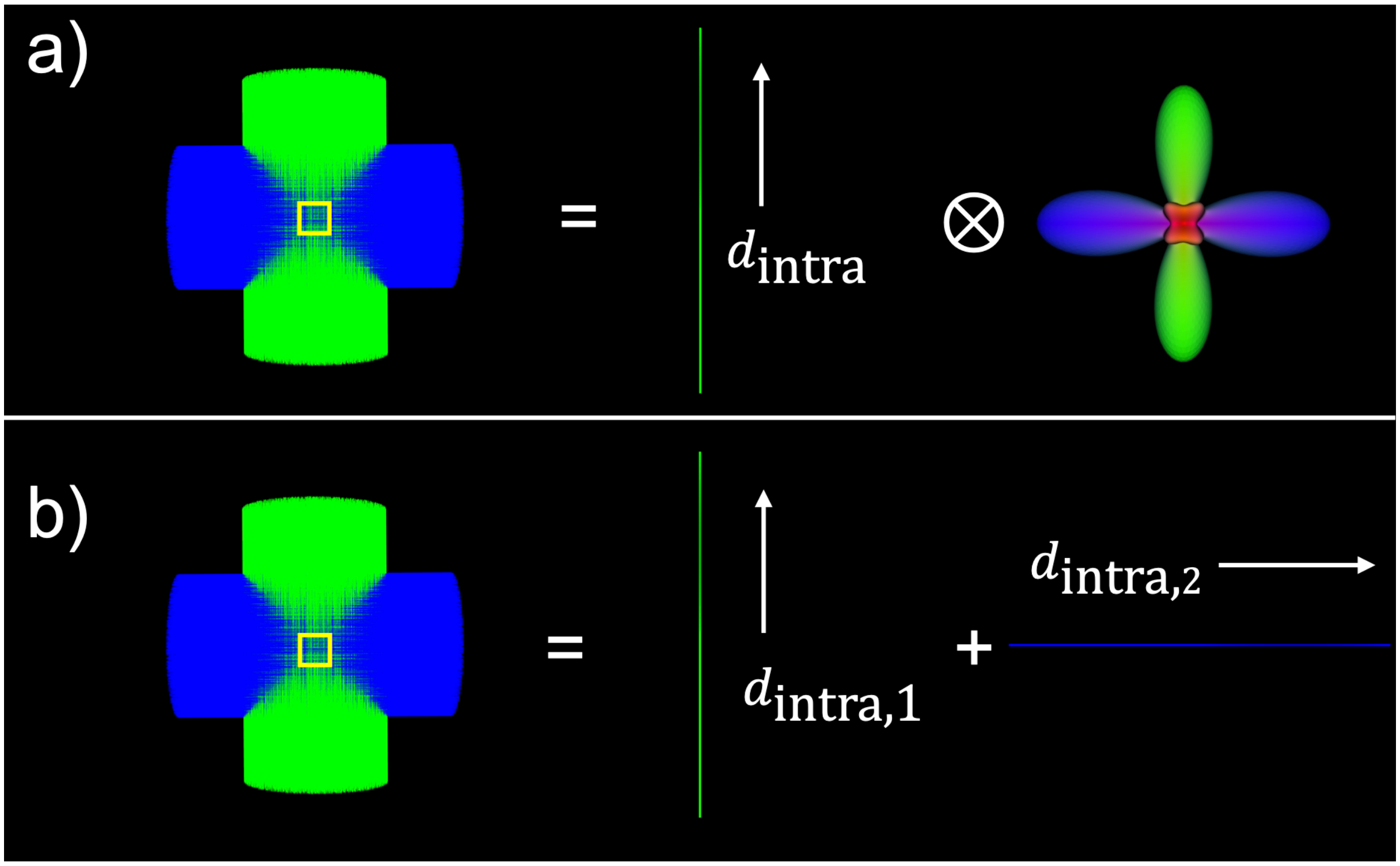

Figure 2. If we focus on just the intra-axonal compartment, the Standard Model of diffusion MRI assumes fiber properties such as intra-axonal diffusivities are consistent within a voxel, and the signal is equal to this fiber response convolved with an FOD (a). Here, instead of a single intra-axonal component dispersed according to a fiber orientation distribution, we assume a set number of fibers with individual diffusivities that vary between fibers (b).

## 2. METHOD

### Theory

The learning objective for fitting microstructural parameters involves maximizing the likelihood of a set of parameters $\theta$ given an observed diffusion signal $S_{\text{obs}}$, i.e.,

$$\arg\max_{\theta} \mathcal{L}(\theta\,|S_{\text{obs}}) = \arg\max_{\theta} P(S_{\text{obs}}|\theta)\,. \tag{1}$$

This framework assumes a specific model with a set of parameters $\theta$. For the Standard Model, these parameters are the volume fractions and diffusivities of the intra-axonal, extra-axonal, and free water compartments along with the orientations of the fibers in the intra-axonal compartment, typically represented using a spherical harmonic basis. Here, we focus on a multi-stick model, which assumes that the extra-axonal and free water compartments have a volume fraction of zero and that the fibers are oriented along a single direction. We chose a multi-stick model because it is the simplest model that describes both fiber directions and fiber properties, though we could choose any microstructural model for this learning framework. Therefore, the diffusion signal $S$ at a b-value $b$ and gradient direction $\vec{g}$ is

$$S(b,\vec{g}) = S_0 \sum_{k=1}^{K} f_k \exp(-b d_{\text{intra,k}}(\vec{n}_k\ \cdot \vec{g})^2)\,, \tag{2}$$

where there are $K$ fibers along directions $\vec{n}_k$ with volume fractions $f_k$ and diffusivities $d_{\text{intra},k}$, and $S_0$ is the signal at $b = 0$ s/mm$^2$.

Using the learning objective in Eq. 1 requires us to assume the number of fibers in a voxel, but this number is not known a priori. Instead, we want to fit a set of alternative candidate models with different numbers of fibers and weight them by prior knowledge to regularize the model selection. We can obtain this fitting objective by marginalizing the probability across the set of models $M_k$ for $k = \{1,2,3\}$:

$$\arg\max_\theta P(S_{\text{obs}}|\theta) = \arg\max_\theta \sum_{k=1}^{K} P(S_{\text{obs}}|\theta_k, M_k) P(M_k)\,. \tag{3}$$

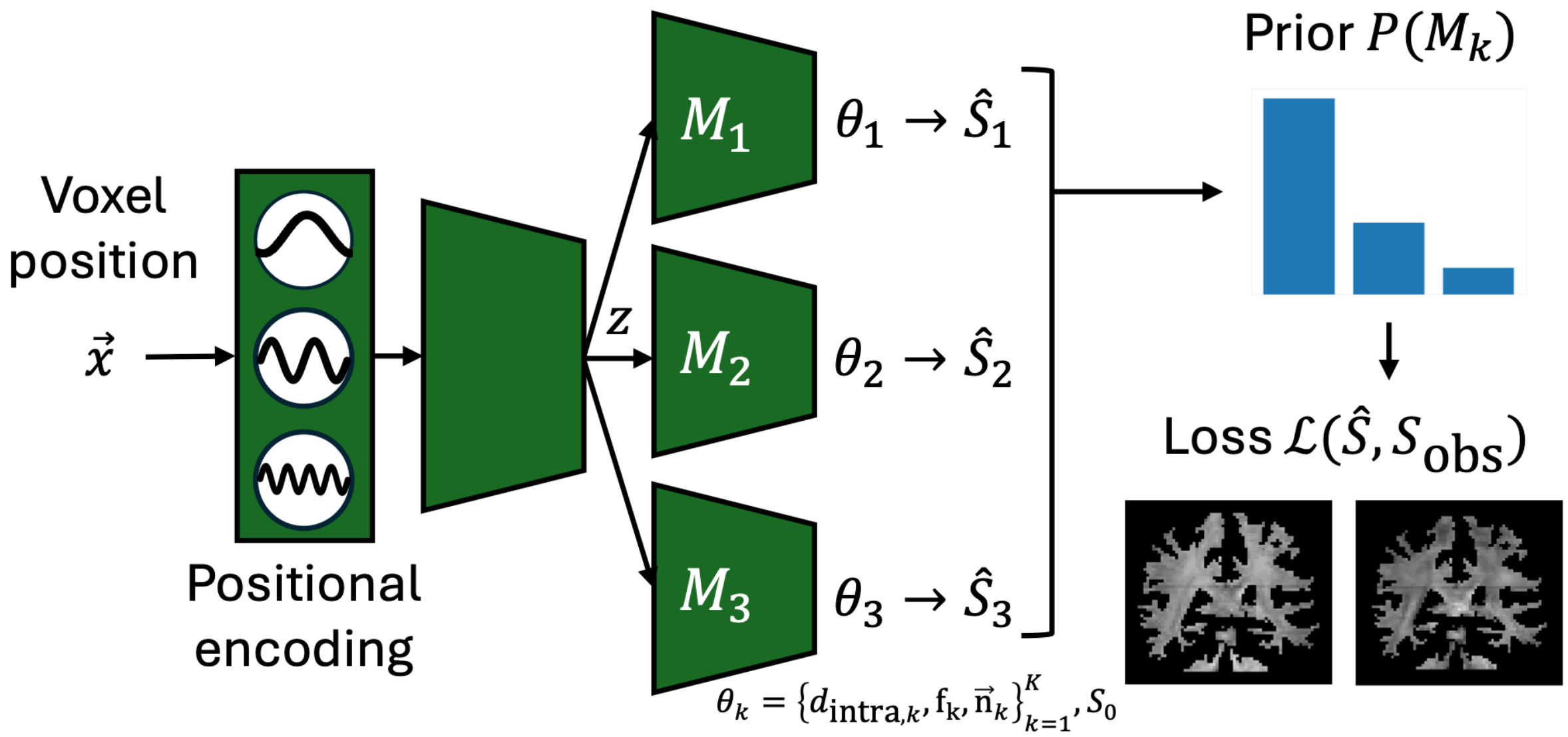

Figure 3. We use an implicit neural field called FiberField, which estimates a set of microstructural parameters according to several candidate models (i.e., 1, 2, or 3 white matter fibers). We calculate the diffusion signal under each model and combine this with prior knowledge to produce a complexity-weighted loss.

To encourage simpler models, we assume the prior on the number of fibers is $P(M_k) = C\exp(-\alpha k)$, where $\alpha$ is a tunable parameter that we set to $\alpha = 0.002$ and $C = 1/\sum_{k=1}^{K}\exp(-\alpha k)$ is a constant parameter so that our prior integrates to 1. This prior is the maximum entropy prior for a non-negative variable with a fixed mean, that is, the prior distribution that assumes the least information[21]. We chose $\alpha = 0.002$ from a parameter sweep as it resulted in the most accurate detection of simulated fiber crossings at 20° and 30° angular separation (see below for simulation details). Maximizing this learning objective is similar to Bayesian model averaging[21], but we restrict the set of models based on number of fibers instead of considering all possible models. If we did not incorporate this prior, the model with the most fibers would always have the lowest loss since it has more parameters to fit the data.

We could optimize the objective in Eq. 3 using any optimization framework. Here, inspired by previous work by Hendriks et al.[19,20], we choose to estimate microstructural parameters across the coordinates of an image using an implicit neural field, named FiberField. An implicit neural field is a neural network that approximates a function over a coordinate system, popularized for their use in scene estimation[22]. Using an implicit neural field incorporates spatial context from nearby voxels, unlike traditional fitting methods that fit each voxel independently. The input to the model is the voxel position, normalized to the range $[-1, 1]$ in each dimension. We add a positional encoding $p(\vec{x}) = [\cos(2\pi A\vec{x}), \sin(2\pi A\vec{x})]$ where $A$ is a random matrix of $n_p \times 3$ normally distributed values with standard deviation $\sigma$. Next, we use a shallow multilayer perceptron to produce a latent vector $z$ of size $n_h$. From $z$, we use a separate head to estimate each microstructural parameter for each of the models $M_k$, that is, the intra-axonal diffusivities $d_{intra,k}$ (normalized to the range $[0, 4 \times 10^{-3}$ mm$^2$/s]), the volume fractions $f_k$ (normalized to sum to 1), the directions of the fibers $\vec{n}_k$, and the signal $S_0$ at $b = 0$ s/mm$^2$ (normalized to be non-negative), as shown in Fig. 3. Based on Hendriks et al.[20], we set the hyperparameters to $\sigma = 4, n_p = 5000$, and $n_h = 2048$. For each candidate model $M_k$, we calculate the signal $\hat{S}_k$ using Eq. 2. Assuming that the residuals of the diffusion signal are normally distributed, the negative log likelihood loss that corresponds to Eq. 3 is

$$\mathcal{L}(S_{\text{obs}}, \hat{S}_k) = -\log\sum_{k=1}^{K}\exp\left(-\left\|S_{\text{obs}} - \hat{S}_k\right\|_2 - \alpha k\right). \tag{4}$$

Table 1. Microstructural parameters used for simulating crossing fibers.

| Name | Bundle 1 value | Bundle 2 value | Value |
|---|---|---|---|
| $d_{\text{intra}}$ (mm$^2$/s) | 0.0018 | 0.0024 | N/A |
| $d_{\text{extra}}^{\|\|}$ (mm$^2$/s) | 0.0012 | 0.0015 | N/A |
| $d_{\text{extra}}^{\perp}$ (mm$^2$/s) | 0.0004 | 0.0007 | N/A |
| $d_{\text{FW}}$ (mm$^2$/s) | N/A | N/A | 0.003 |
| $f_{\text{intra}}$ | 0.425 | 0.425 | N/A |
| $f_{\text{extra}}$ | 0.05 | 0.05 | N/A |
| $f_{\text{FW}}$ | N/A | N/A | 0.05 |

We train the model using the Adam optimizer[23] with a learning rate of $10^{-4}$ for 1000 epochs or until the total loss has not decreased in the last 250 epochs, checking every 10 epochs. Because training is self-supervised, there is no held out validation set. At inference, we filtered out voxels with diffusivity or volume fraction smaller than $10^{-4}$. We trained the model using an NVIDIA RTX 6000 GPU. Code for FiberField is available online in open source at https://github.com/MASILab/FiberField.

**Simulations**

Because we do not have access to ground truth values for fiber directions and diffusivities, we first tested FiberField on simulated diffusion MRI using Fiberfox[24]. Fiberfox is a diffusion MRI simulator that produces diffusion MRI from tractography. We generated 10,000 fibers in cylinders of radius 4 at crossings with angular separations from 5° to 90°. Although we only use FiberField to estimate an intra-axonal compartment, we simulated data with the full Standard Model to more closely align with in vivo data. Table 1 gives the microstructural parameters used in the simulation, chosen such that the bundles have slightly different diffusivities within the range of previously reported values[11]. In order to use unique parameters for each bundle, we simulated each bundle independently, averaged the real and imaginary parts from the simulated acquisitions, and calculated the magnitude. We added Rician noise with an SNR of 60 to produce 10 simulations at each angular separation.

We simulated the diffusion MRI using the same gradient directions and b-values as the Human Connectome Project acquisitions[25], with 270 directions across 3 shells at $b = 1000$, $b = 2000$, and $b = 3000$ s/mm$^2$ and 18 $b = 0$ s/mm$^2$ volumes.

**In vivo acquisition**

We also applied FiberField to diffusion MRI acquired from a human subject using a high angular resolution diffusion imaging (HARDI)-style acquisition. The acquisition had an isotropic spatial resolution of 2 mm, 1 $b = 0$ s/mm$^2$ volume, and 168 directions across 3 shells at $b = 800$, $b = 1600$, and $b = 2400$ s/mm$^2$ along with a T1-weighted image acquired at 1 mm isotropic resolution. The data were deidentified and acquired under a protocol approved by the institutional review board (IRB #020619). We preprocessed the image using PreQual[26], which includes normalization, susceptibility distortion correction with a reverse phase encoding image synthesized from the T1-weighted image with Synb0-DisCo[27], and eddy current and motion correction[28]. We segmented the white matter using a FreeSurfer parcellation[29] generated from the T1-weighted image acquired during the same scanning session registered to the diffusion MRI using ANTs[30].

**Baselines and validation metrics**

We use two methods for comparison. Our first baseline is MRtrix3's fod2fixel[15], which finds fiber directions from fiber orientation distributions (FODs) calculated using constrained spherical deconvolution (CSD)[4]. We calculate FODs at order 8 using a response function estimated from the white matter using Tournier et al.'s method[31], and we normalized the FODs to integrate to 1. Fod2fixel segments each FOD into lobes and finds the weighted peak direction of each lobe, rejecting peaks that are smaller in amplitude than 0.1. While fod2fixel provides metrics for fiber spread and cross-section based on the FOD, these metrics are not directly comparable to those calculated from the Standard Model.

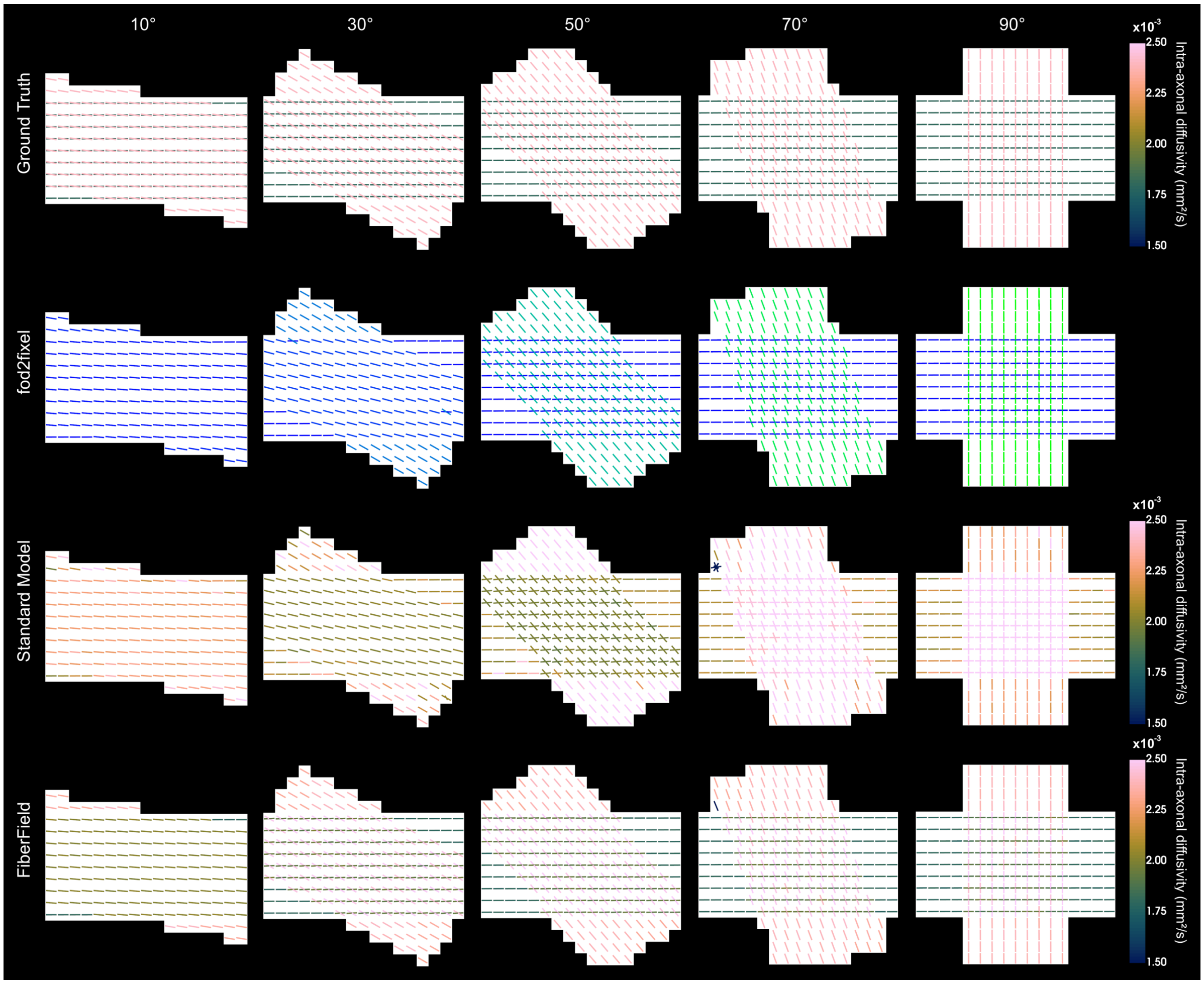

Figure 4. On diffusion MRI data simulated from FiberFox, FiberField can resolve fibers at lower directions and more accurately estimate diffusivities of crossing fibers. All three models predict the mean direction when they are unable to detect the secondary crossing fiber (see crossings at 30°). The Standard Model overestimates diffusivities compared to ground truth, especially at fibers with high angular separations (see the Standard Model estimates at 90°) Note that fod2fixel provides directions with no estimate of diffusivities.

We also used the Standard Model as our second baseline to estimate voxel-level microstructural features using the method of Coelho et al.[11] To map the voxel-level parameters onto fibers, we also used fod2fixel to find the directions of the FOD calculated from Standard Model fitting, and we mapped the Standard Model parameters onto all fibers within a voxel. For fod2fixel and Standard Model fitting, we used a workstation with 20 Intel Xeon W-2255 CPUs.

For comparison in simulation, we matched the fibers estimated for all methods to the ground truth fibers based on fiber direction. We calculated angular error, intra-axonal diffusivity root mean square error (RMSE), and fiber sensitivity to demonstrate how well we detect the fibers in a voxel. For comparison in vivo, we matched the fibers from FiberField and the Standard Model to those from fod2fixel and calculated angular difference between the two, as well as mean square difference between the diffusivities calculated from FiberField and the Standard Model.

In addition to fitting FiberField with a multi-stick model, we also applied FiberField using the full Standard Model (i.e., intra-axonal, extra-axonal, and free water compartments) with discrete fiber directions as the FOD for comparison. For this model, we fixed the diffusivity of free water to $d_{\mathrm{FW}} = 0.003$ mm$^2$/s.

## 3. RESULTS

### Simulations

Figure 4 visualizes the diffusivities and fiber directions estimated from each method. The directions estimated from fod2fixel from FODs calculated using CSD and fod2fixel from FODs calculated using the Standard Model were similar. At 30° angular separation, FiberField detected secondary fibers that fod2fixel did not. All models estimated the mean direction when they could not distinguish fibers. The Standard Model tended to overestimate the intra-axonal diffusivity, especially at fiber crossings with high angular separation.

Looking at the quantitative error in simulation (Fig. 5), the diffusivity RMSE and angular error were high at low angular separations where the models were not able to distinguish both fibers. FiberField had a 25% increase in angular sensitivity compared to fod2fixel. It also had a lower angular error and intra-axonal diffusivity RMSE compared to the other methods. We compared the angular error and diffusivity RMSE at each value of angular separation using a paired $t$-test with Benjamini-Hochberg multiple comparisons correction ($p < 0.05$). The difference in angular error was significant between FiberField and fod2fixel as well as FiberField and the Standard Model at all angular separations above 10°. The difference in diffusivity RMSE was also significant between FiberField and the Standard Model for all separations above 10°. The diffusivity RMSE for FiberField is 52% smaller than the Standard Model within fiber crossings.

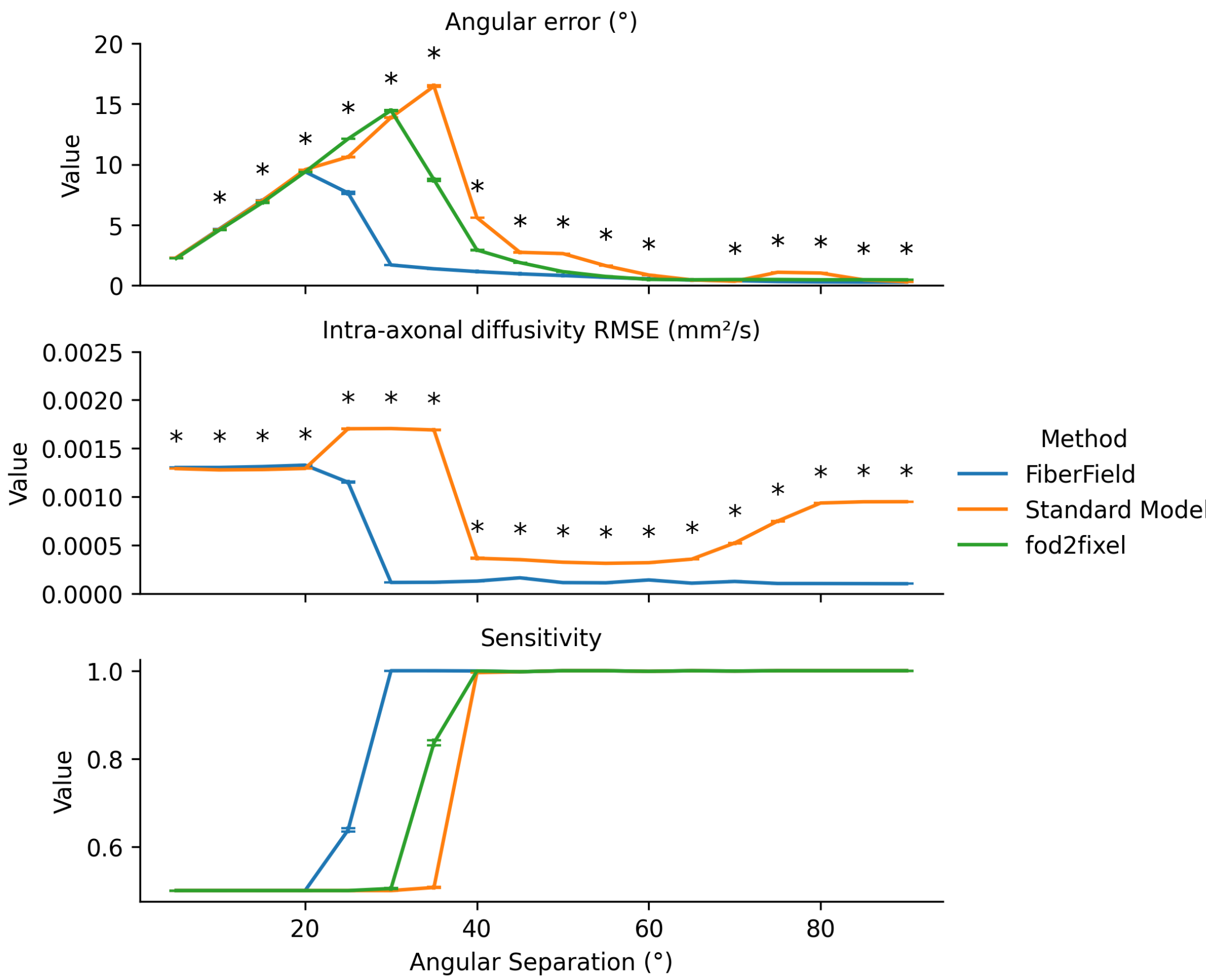

Figure 5. FiberField accurately estimates fiber diffusivities and directions up to about 30° angular separation in simulation, compared to about 40° for fod2fixel and Standard Model with fod2fixel. FiberField is significantly more accurate than the Standard Model and fod2fixel in terms of angular error and intra-axonal diffusivity RMSE (*: $p < 0.05$ between FiberField and both methods, paired $t$-test with Benjamini-Hochberg multiple comparisons correction). None of the methods are sensitive to noise at an SNR of 60 (error bars indicate 95% confidence intervals).

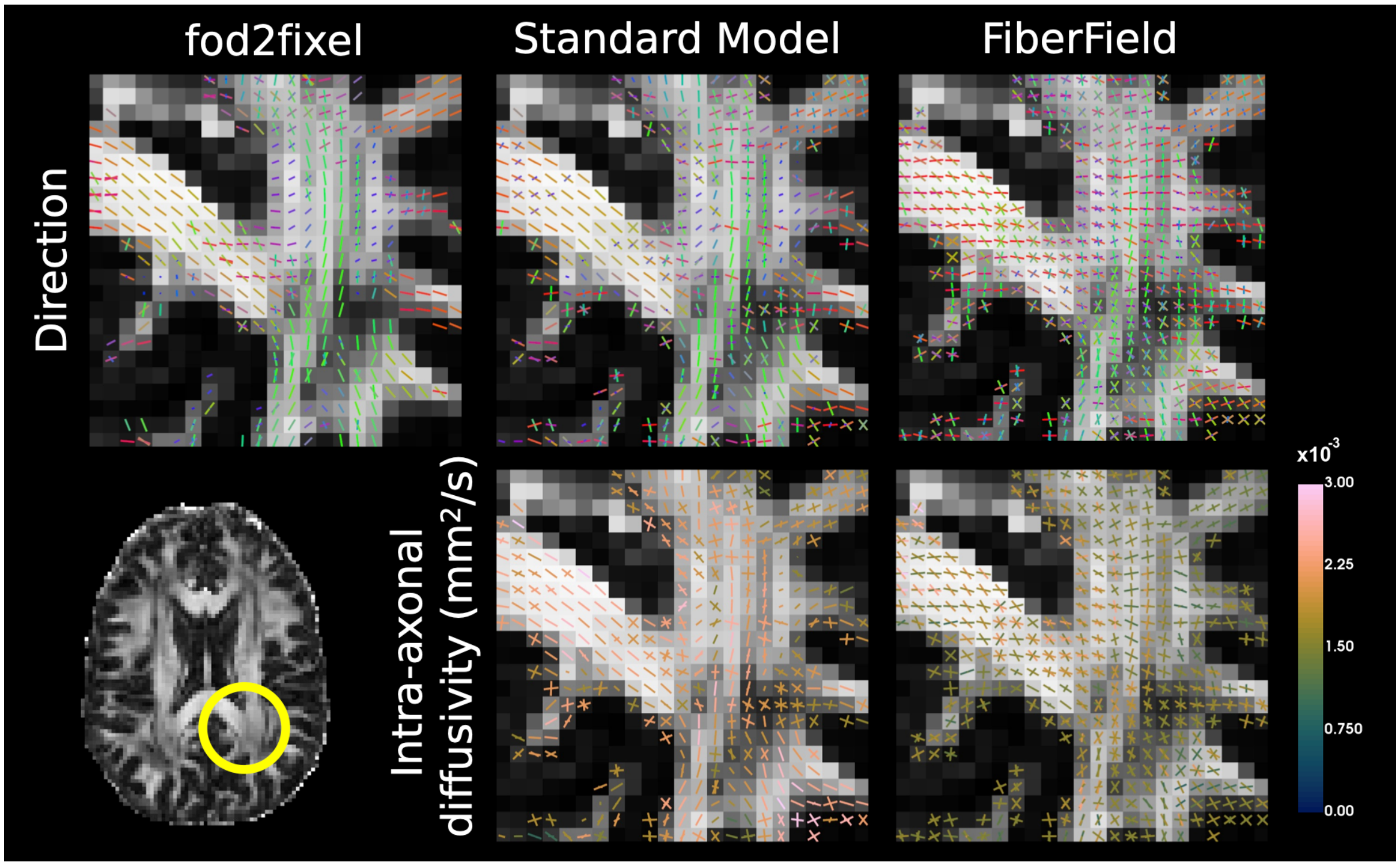

Figure 6. On acquired HARDI diffusion MRI, the fiber directions estimated from FiberField are similar to those estimated from MRtrix3's fod2fixel, though FiberField overestimates the number of fibers compared to fod2fixel in the corpus callosum. Standard Model fitting results in higher diffusivity estimates that our FiberField fitting, consistent with results in simulation.

**In vivo acquisition**

Next, we applied each method to a single HARDI-style acquisition on data acquired in vivo from a human subject. Figure 6 visualizes a region of interest near the intersection of the corpus callosum and internal capsule. The directions identified from each method were similar, although FiberField tended to overestimate the number of fibers relative to the other two methods. For example, the corpus callosum contained two to three fibers in most voxels using FiberField, while the other methods estimated a single fiber in most voxels in this region.

The diffusivity estimates from FiberField were lower than those estimated from the Standard Model. This result is consistent with the simulation, where the Standard Model overestimated the intra-axonal diffusivity relative to ground truth. The overall angular error on average was $2.2 \pm 2.8°$, the intra-axonal diffusivity RMSE was $4.4 \times 10^{-4} \pm 5.2 \times 10^{-4}$ $mm^2/s$, and the fiber sensitivity was $0.87 \pm 0.21$ using FiberField.

Training and inference for FiberField lasted 29 minutes and 44 seconds for the in vivo acquisition on an NVIDIA RTX 6000 GPU, while fod2fixel was 17 seconds and Standard Model fitting was 5 minutes and 37 seconds.

**Fitting the full Standard Model**

Finally, we applied FiberField on simulated data to estimate the full Standard Model parameters, including an extra-axonal compartment and a free water compartment. Figure 7 visualizes Standard Model parameter estimates for the simulated fibers. FiberField was more accurate than the Standard Model in the crossing fiber region, where the Standard Model overestimated diffusivities and underestimated the free water fraction. However, outside of the fiber crossings, FiberField identified several nonexistent fibers. These nonexistent fibers had small volume fractions and small diffusivities, and they aligned in direction with the second bundle despite being outside the region of the second bundle.

After matching the fibers identified from FiberField to those identified from fod2fixel, the average angular difference between the two methods was 10.5 ± 7.1° in the white matter.

## 4. DISCUSSION

FiberField provides a similar modeling framework to models such as COMMIT and FORCE that model fiber direction and microstructural properties together. Like FiberField, the authors of COMMIT found that voxelwise models such as NODDI fail in regions with fiber crossings.[3] However, FiberField differs from COMMIT in that it does not require tractography to be generated ahead of time. FORCE does not assume a single model, instead opting to provide many metrics from different modeling frameworks such as diffusion tensor imaging, diffusion kurtosis imaging, NODDI, etc. Novikov et al. discussed how all the different microstructural models are in reality different mathematical representations that do not jointly represent new information[32]. With FiberField, we opt to limit the set of candidate models to the relevant degree of freedom between voxels (the number of fibers in a voxel), instead of choosing from the vast potential set of microstructural models. The actual microstructural model backbone of FiberField can be any microstructural model, but we demonstrated a multi-stick model and the Standard Model here. While we did not test what diffusion acquisition is sufficient for fitting FiberField, Coelho et al. optimized acquisition parameters for Standard Model fitting[11]. By changing the microstructural model backbone, FiberField could be tailored to specific protocols and acquisition schemes depending on available data or scanning time constraints.

FiberField offers a method for simultaneous estimation of microstructural parameters and selection among alternative candidate models with different numbers of fibers using prior information. We presented a preliminary analysis focusing on simulated data because it allows for ground truth comparisons, but we can consider several potential applications of this model for continued exploration. Because it provides fiber-specific metrics, FiberField could power analyses similar to fixel-based analysis, where we aim to map fiber metrics into a common spatial framework among subjects[15]. FiberField could also allow for tractometry, mapping the spatial changes in metrics along the length of a bundle[33]. In addition, the multi-stick model we estimated here could be treated as fiber-specific response functions that could be used for fiber-specific constrained spherical deconvolution[34].

In simulation, FiberField outperformed fod2fixel in angular error, and it outperformed the Standard Model in diffusivity estimation in fiber crossings. For this work, we selected $P(M_k) = C\,\exp(-\alpha k)$ as the prior for the number of fibers for our set of candidate models. While we chose $\alpha = 0.002$ based on a parameter sweep on simulated crossings, we found that FiberField estimated two to three fibers in the corpus callosum, a region where we expect a single fiber. Potentially, there was some leakage in the spatial representation, as the spatial representation encourages smooth properties despite boundaries at different white matter tracts. It is also possible that this value for $\alpha$ and/or this prior is too liberal for our in vivo acquisition. The value represents a tradeoff between fiber specificity (i.e., too high will overestimate the number of fibers) and fiber sensitivity (too low will not allow for crossings at small angular separations). Further work is needed to justify this prior, and ideally, the prior should be dependent on the noise level or perhaps learned directly from the diffusion MRI. While we showed that FiberField can be applied in vivo, the model needs further validation, especially to see if microstructural parameter estimates are more consistent along tracts.

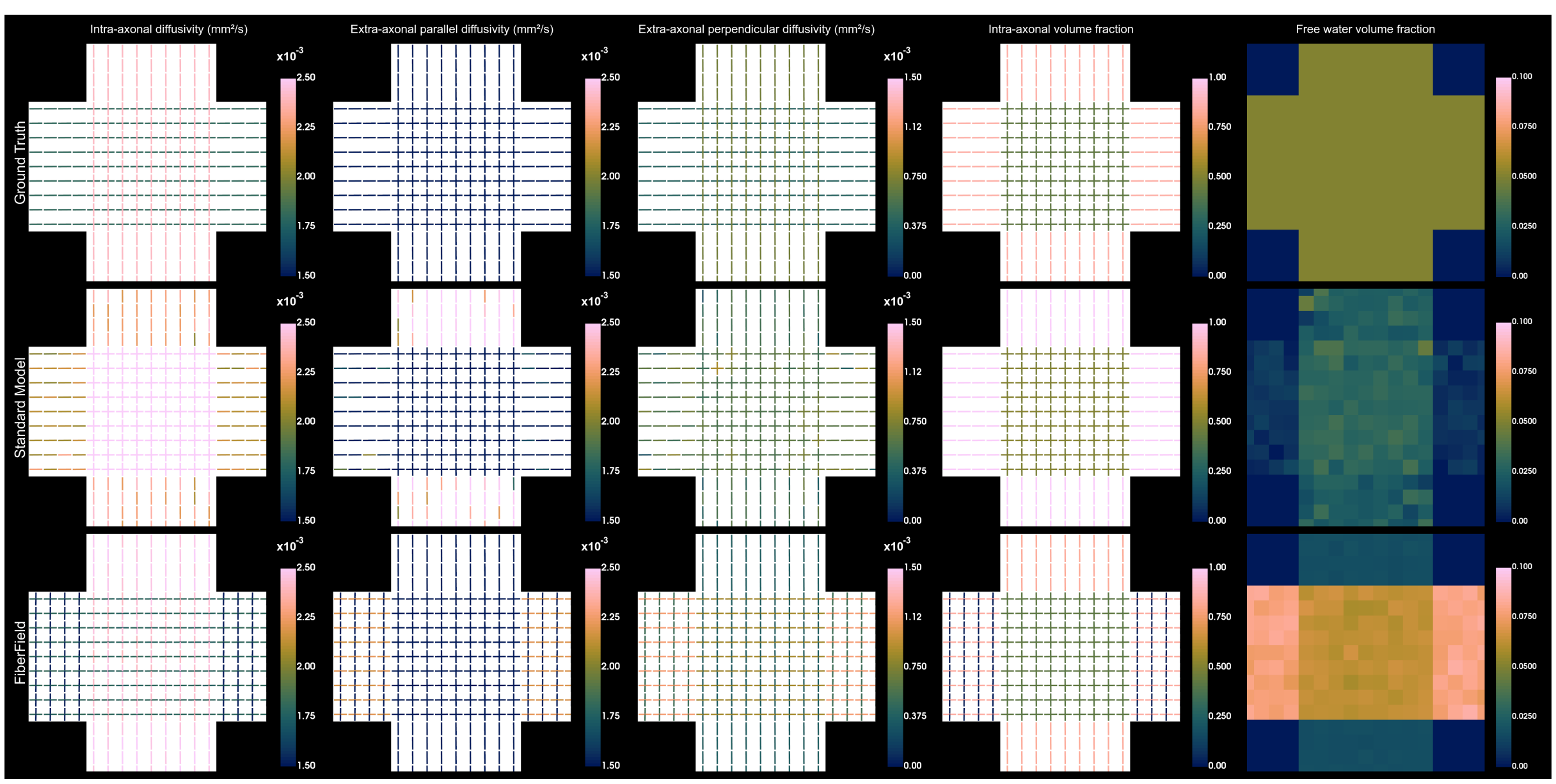

Figure 7. We can also fit the full Standard Model parameters using FiberField. FiberField is more accurate than typical Standard Model fitting at estimating diffusivities and free water volume fractions within fiber crossings, but outside of the crossings, FiberField identifies non-existent fibers with high diffusivities and low volume fractions.

## 5. CONCLUSION

We presented FiberField, a method for modeling fiber directions and properties using implicit neural fields. Using a framework inspired by Bayesian model averaging, we introduced a model selection loss that allows for jointly optimizing and selecting from several candidate models based on the number of fibers within a voxel. Our model outperforms traditional voxel-wise modeling in terms of angular error and intra-axonal diffusivity estimation in simulation, and we demonstrated its use on in vivo data. FiberField allows for adaptive modeling of white matter with joint estimation of microstructural properties and fiber directions.

## ACKNOWLEDGEMENTS

This work was funded by NIH 1R01NS136743-01A1, NIH 1R01EB017230, and NIH P50HD103537 (VKC). Data were provided in part by the Human Connectome Project, WU-Minn Consortium (Principal Investigators: David Van Essen and Kamil Ugurbil; 1U54MH091657) funded by the 16 NIH Institutes and Centers that support the NIH Blueprint for Neuroscience Research; and by the McDonnell Center for Systems Neuroscience at Washington University. This work was supported by the Alzheimer's Disease Sequencing Project Phenotype Harmonization Consortium (ADSP-PHC) that is funded by NIA (U24 AG074855, U01 AG068057 and R01 AG059716). We used generative AI to create code segments based on task descriptions, as well as to debug, edit, and autocomplete code. Additionally, generative AI technologies have been employed to assist in structuring sentences and performing grammatical checks. We take accountability for the review of all content generated by AI in this work.

## REFERENCES

[1] Kim, M. E., Gao, C., Ramadass, K., Newlin, N. R., Kanakaraj, P., Bogdanov, S., Rudravaram, G., Archer, D., Hohman, T. J., Jefferson, A. L., Morgan, V. L., Roche, A., Englot, D. J., Resnick, S. M., Beason-Held, L. L., Cutting, L. E., Barquero, L. A., D'archangel, M. A., Nguyen, T. Q., et al., "White matter micro- and macrostructure brain charts for the human lifespan," Nature, 1–11 (2026).

[2] Jeurissen, B., Descoteaux, M., Mori, S. and Leemans, A., "Diffusion MRI fiber tractography of the brain," NMR in Biomedicine **32**(4) (2019).

[3] Daducci, A., Dal Palù, A., Lemkaddem, A. and Thiran, J.-P., "COMMIT: Convex Optimization Modeling for Microstructure Informed Tractography," IEEE Transactions on Medical Imaging **34**(1), 246–257 (2015).

[4] Tournier, J.-D., Calamante, F. and Connelly, A., "Robust determination of the fibre orientation distribution in diffusion MRI: Non-negativity constrained super-resolved spherical deconvolution," NeuroImage **35**(4), 1459–1472 (2007).

[5] Jeurissen, B., Leemans, A., Tournier, J., Jones, D. K. and Sijbers, J., "Investigating the prevalence of complex fiber configurations in white matter tissue with diffusion magnetic resonance imaging," Human Brain Mapping **34**(11), 2747–2766 (2013).

[6] McMaster, E. M., Newlin, N., Schwartz, T., Saunders, A. M., Rudravaram, G., Suh, Y., Yoon, J., Kim, M. E., Cho, C., Ramadass, K., Liu, Y., Zuo, L., Garyfallidis, E., Nir, T. M., Jahanshad, N., Schilling, K., Moyer, D. and Landman, B. A., "Introducing Beyond FA: Crowdsourcing Imaging Biomarkers for Alzheimer's Disease," presented at Medical Imaging with Deep Learning - Short Papers, 12 April 2025.

[7] Yang, J., Poot, D. H. J., Caan, M. W. A., Su, T., Majoie, C. B. L. M., van Vliet, L. J. and Vos, F. M., "Reliable Dual Tensor Model Estimation in Single and Crossing Fibers Based on Jeffreys Prior," PLOS ONE **11**(10), e0164336 (2016).
[8] Caan, M. W. A., Khedoe, H. G., Poot, D. H. J., den Dekker, A. J., Olabarriaga, S. D., Grimbergen, K. A., van Vliet, L. J. and Vos, F. M., "Estimation of Diffusion Properties in Crossing Fiber Bundles," IEEE Transactions on Medical Imaging **29**(8), 1504–1515 (2010).
[9] Jeurissen, B., Tournier, J. D., Dhollander, T., Connelly, A. and Sijbers, J., "Multi-tissue constrained spherical deconvolution for improved analysis of multi-shell diffusion MRI data," NeuroImage **103**, 411–426 (2014).
[10] Anderson, A. W., "Measurement of fiber orientation distributions using high angular resolution diffusion imaging," Magnetic Resonance in Medicine **54**(5), 1194–1206 (2005).
[11] Coelho, S., Baete, S. H., Lemberskiy, G., Ades-Aron, B., Barrol, G., Veraart, J., Novikov, D. S. and Fieremans, E., "Reproducibility of the Standard Model of diffusion in white matter on clinical MRI systems," NeuroImage **257**, 119290 (2022).
[12] Farquharson, S., Tournier, J.-D., Calamante, F., Fabinyi, G., Schneider-Kolsky, M., Jackson, G. D. and Connelly, A., "White matter fiber tractography: why we need to move beyond DTI: Clinical article," Journal of Neurosurgery **118**(6), 1367–1377 (2013).
[13] Tournier, J.-D., Calamante, F., Gadian, D. G. and Connelly, A., "Direct estimation of the fiber orientation density function from diffusion-weighted MRI data using spherical deconvolution," NeuroImage **23**(3), 1176–1185 (2004).
[14] Smith, R. E., Tournier, J. D., Calamante, F. and Connelly, A., "SIFT: Spherical-deconvolution informed filtering of tractograms," NeuroImage **67**, 298–312 (2013).
[15] Dhollander, T., Clemente, A., Singh, M., Boonstra, F., Civier, O., Duque, J. D., Egorova, N., Enticott, P., Fuelscher, I., Gajamange, S., Genc, S., Gottlieb, E., Hyde, C., Imms, P., Kelly, C., Kirkovski, M., Kolbe, S., Liang, X., Malhotra, A., et al., "Fixel-based Analysis of Diffusion MRI: Methods, Applications, Challenges and Opportunities," NeuroImage **241**, 118417 (2021).
[16] Saunders, A. M., Remedios, L. W., McMaster, E. M., Yoon, J., Rudravaram, G., Sadriddinov, A., Kanakaraj, P., Landman, B. A. and Anderson, A. W., "DeepFixel: crossing white-matter fiber identification through spherical convolutional neural networks," Medical Imaging 2026: Clinical and Biomedical Imaging **13929**, 618–628, SPIE (2026).
[17] Schilling, K. G., Gao, Y., Stepniewska, I., Janve, V., Landman, B. A. and Anderson, A. W., "Histologically derived fiber response functions for diffusion MRI vary across white matter fibers—An ex vivo validation study in the squirrel monkey brain," NMR in Biomedicine **32**(6), e4090 (2019).
[18] Shah, A. J., Henriques, R. N., Ramirez-Manzanares, A., Filipiak, P., Baete, S., Deka, K., Gor, M., Koudoro, S. and Garyfallidis, E., "FORCE: FORward modeling for Complex microstructure Estimation" (2025).
[19] Hendriks, T., Vilanova, A. and Chamberland, M., "Implicit neural representation of multi-shell constrained spherical deconvolution for continuous modeling of diffusion MRI," Imaging Neuroscience **3** (2025).
[20] Hendriks, T., Arends, G., Versteeg, E., Vilanova, A., Chamberland, M. and Tax, C. M. W., "Implicit neural representations for accurate estimation of the standard model of white matter" (2025).
[21] Hoeting, J. A., Madigan, D., Raftery, A. E. and Volinsky, C. T., "Bayesian Model Averaging: A Tutorial."
[22] Mildenhall, B., Srinivasan, P. P., Tancik, M., Barron, J. T., Ramamoorthi, R. and Ng, R., "NeRF: Representing Scenes as Neural Radiance Fields for View Synthesis," Computer Vision – ECCV 2020, A. Vedaldi, H. Bischof, T. Brox, and J.-M. Frahm, Eds., 405–421, Springer International Publishing, Cham (2020).
[23] Kingma, D. P. and Ba, J., "Adam: A Method for Stochastic Optimization," 3rd International Conference for Learning Representations (2014).
[24] Neher, P. F., Laun, F. B., Stieltjes, B. and Maier-Hein, K. H., "Fiberfox: Facilitating the creation of realistic white matter software phantoms," Magnetic Resonance in Medicine **72**(5), 1460–1470 (2014).
[25] Sotiropoulos, S. N., Jbabdi, S., Xu, J., Andersson, J. L., Moeller, S., Auerbach, E. J., Glasser, M. F., Hernandez, M., Sapiro, G., Jenkinson, M., Feinberg, D. A., Yacoub, E., Lenglet, C., Van Essen, D. C., Ugurbil, K. and Behrens, T. E. J., "Advances in diffusion MRI acquisition and processing in the Human Connectome Project," NeuroImage **80**, 125–143 (2013).
[26] Cai, L. Y., Yang, Q., Hansen, C. B., Nath, V., Ramadass, K., Johnson, G. W., Conrad, B. N., Boyd, B. D., Begnoche, J. P., Beason-Held, L. L., Shafer, A. T., Resnick, S. M., Taylor, W. D., Price, G. R., Morgan, V. L., Rogers, B. P., Schilling, K. G. and Landman, B. A., "PreQual: An automated pipeline for integrated preprocessing

and quality assurance of diffusion weighted MRI images," Magnetic Resonance in Medicine **86**(1), 456–470 (2021).
[27] Schilling, K. G., Blaber, J., Huo, Y., Newton, A., Hansen, C., Nath, V., Shafer, A. T., Williams, O., Resnick, S. M., Rogers, B., Anderson, A. W. and Landman, B. A., "Synthesized b0 for diffusion distortion correction (Synb0-DisCo)," Magnetic Resonance Imaging **64**, 62–70 (2019).
[28] Andersson, J. L. R. and Sotiropoulos, S. N., "An integrated approach to correction for off-resonance effects and subject movement in diffusion MR imaging," NeuroImage **125**, 1063–1078 (2016).
[29] Fischl, B., "FreeSurfer," NeuroImage **62**(2), 774–781 (2012).
[30] Avants, B. B., Tustison, N. J., Song, G., Cook, P. A., Klein, A. and Gee, J. C., "A Reproducible Evaluation of ANTs Similarity Metric Performance in Brain Image Registration," NeuroImage **54**(3), 2033 (2011).
[31] Tournier, J.-D., Calamante, F. and Connelly, A., "Determination of the appropriate b value and number of gradient directions for high-angular-resolution diffusion-weighted imaging," NMR in Biomedicine **26**(12), 1775–1786 (2013).
[32] Novikov, D. S., Kiselev, V. G. and Jespersen, S. N., "On modeling," Magnetic Resonance in Medicine **79**(6), 3172–3193 (2018).
[33] Bells, S., Cercignani, M., Deoni, S., Assaf, Y., Pasternak, O., Evans, C. J., Leemans, A. and Jones, D. K., "Tractometry – Comprehensive Multi-modal Quantitative Assessment of White Matter Along Specific Tracts."
[34] Dhollander, T., Raffelt, D. and Connelly, A., "Unsupervised 3-tissue response function estimation from single-shell or multi-shell diffusion MR data without a co-registered T1 image," ISMRM Workshop on Breaking the Barriers of Diffusion MRI (2016).